\documentclass[review]{elsarticle}

\usepackage{hyperref}
\usepackage{graphicx}
\usepackage{amssymb}
\usepackage{amsthm}
\usepackage{amsmath}
\usepackage{lineno}
\usepackage{natbib}
\usepackage{bm}
\usepackage{float}
\usepackage{multirow}
\usepackage{changepage}
\usepackage{pdflscape}
\usepackage{comment}

\newcommand{\etal}{\textit{et al.}}
\newcommand{\logit}{\text{logit}}
\newcommand{\expit}{\text{expit}}
\newcommand\numberthis{\addtocounter{equation}{1}\tag{\theequation}}

\journal{ }

\begin{document}

\begin{frontmatter}

\title{Modeling and presentation of vaccination coverage estimates using data from household surveys} 

\author[biostat]{Tracy Qi Dong\corref{corauthor}}
\ead{qd8@uw.edu}
\cortext[corauthor]{Corresponding author}
\author[biostat,stat]{Jon Wakefield}

\address[biostat]{Department of Biostatistics, University of Washington, Health Sciences Building, NE Pacific St, Seattle, WA 98195, USA}

\address[stat]{Department of Statistics, University of Washington, Padelford Hall, NE Stevens Way, Seattle, WA 98195, USA}

\begin{abstract}

It is becoming increasingly popular to produce high-resolution maps of vaccination coverage by fitting Bayesian geostatistical models to data from household surveys. Often, the surveys adopt a stratified cluster sampling design. We discuss a number of crucial choices with respect to two key aspects of the map production process: the acknowledgement of the survey design in modeling, and the appropriate presentation of estimates and their uncertainties. Specifically, we consider the importance of accounting for survey stratification and cluster-level non-spatial excess variation in survey outcomes when fitting geostatistical models. We also discuss the trade-off between the geographical scale and precision of model-based estimates, and demonstrate visualization methods for mapping and ranking that emphasize the probabilistic interpretation of results. A novel approach to coverage map presentation is proposed to allow comparison and control of the overall map uncertainty level. We use measles vaccination coverage in Nigeria as a motivating example and illustrate the different issues using data from the 2018 Nigeria Demographic and Health Survey.

\end{abstract}


\end{frontmatter}

\section*{Highlights}

\begin{itemize}
\item Acknowledging survey design in geostatistical modeling provides valid inference and improves vaccination coverage estimation.

\item High-resolution coverage maps based on survey data are often associated with large uncertainties. 

\item Visualizing posterior distributions of coverage estimates and rankings help reveal uncertainties and aid interpretation. 

\item A novel presentation approach is proposed to compare and control the overall map precision.
\end{itemize}


\section{Introduction} \label{sec_intro}

There has been an explosion in high-resolution map production for health and demographic indicators \cite{gething2015creating, gething2016mapping, golding2017mapping, steele2017mapping, bosco2017exploring, osgood2018mapping, graetz2018mapping}, including childhood vaccination coverage \cite{takahashi2017geography, utazi2018high, utazi2018spatial, utazi2019mapping, mosser2019mapping, utazi2020geospatial}. The WorldPop project \cite{worldpop} and the Institute for Health Metrics and Evaluation (IHME) \cite{IHME} are two major producers of vaccination coverage surfaces at fine spatial scales. Their estimates are used by researchers and policy makers from organizations across the globe \cite{worldpopuse, IHMEuse}. 

Household surveys, such as the Demographic and Health Surveys (DHS)\cite{DHS} and the Multiple Indicator Cluster Surveys (MICS)\cite{MICS}, are a main data source for mapping vaccination coverage \cite{utazi2018high, utazi2018spatial, utazi2019mapping, mosser2019mapping}. In recent DHS surveys, the groupings of households, known as clusters, are geo-referenced via their Global Positioning System (GPS) coordinates. Both WorldPop and IHME use model-based geostatistics \cite{diggle2019model} to analyze cluster-level survey data. The core of their methods involves the following steps: match survey cluster locations to a collection of geospatial covariates, fit Bayesian spatial regression models to the cluster-level data using a continuous Gaussian Process (GP) model, and produce pixel-level vaccination coverage estimates on a fine grid. Often, the pixel-level estimates are aggregated by population density to form administrative-area-level estimates which are usually more relevant in the context of immunization program monitoring and intervention planning.

This map production framework has a number of advantages. It allows for utilization of geospatial covariates from different sources, avoids the arbitrariness of the definition of neighbors in a discrete spatial model, allows combination of data with different geographical resolutions \cite{wilson2018pointless}, and allows coverage estimation for small administrative areas in which there are few or no survey clusters. However, it also has many challenges that need to be carefully addressed. First, the survey cluster locations are often randomly displaced within a certain radius (known as jittering) to preserve respondent confidentiality \cite{DHSGPS}. This makes it difficult to accurately match the survey clusters to geospatial covariates as the exact cluster locations are unknown. In addition, household surveys in low- and middle-income countries (LMIC) use multi-stage stratified cluster designs, with stratification by geographical region crossed with urban/rural. Not accounting for the survey design is well known to lead to biased estimates when stratification is ignored and anticonservative uncertainty intervals when clustering is not acknowledged \cite{paige2019design}. The appropriate acknowledgement of survey design in a model-based framework is challenging. Last, but not least, most household surveys are powered to provide reliable estimates at a particular sub-national scale (e.g., administrative-1 areas which are one below the national level). This implies that pixel-level vaccination coverage estimates are often associated with high uncertainties due to the sparsity of survey data, and care needs to be taken to appropriately present the estimates with their uncertainties. 

In this paper, we focus on two key aspects of the map production process: the acknowledgement of the survey design in modeling, and the appropriate presentation of estimates and their uncertainties. We use the coverage of the first dose of measles-containing-vaccine (MCV1) among children aged 12--23 months in Nigeria as a motivating example and illustrate modeling and presentation using data from the 2018 Nigeria Demographic and Health Survey (NDHS) \cite{dhs2018}. In Section \ref{sec_2018dhs}, we briefly introduce the 2018 NDHS and present descriptive summary of the data. We then compare common models in Section \ref{sec_mod}, including the model used in Utazi \etal~(2018) \cite{utazi2018high}, to illustrate approaches to modeling survey stratification and cluster-level non-spatial variation. A discussion of the trade-off between the geographical scale and precision of model-based estimates is presented in Section \ref{sec_pres}, along with a demonstration of visualization methods for mapping and ranking that emphasize the probabilistic interpretation of results. We also propose a novel approach to coverage map presentation that allows comparison and control of the overall map uncertainty level. In Section \ref{sec_disc}, we conclude with a general discussion and guidelines for map production and presentation. 


\section{Motivating example: the 2018 Nigeria DHS} \label{sec_2018dhs}

\begin{figure}[H]  
	\centering
	\includegraphics[width=0.6\textwidth]{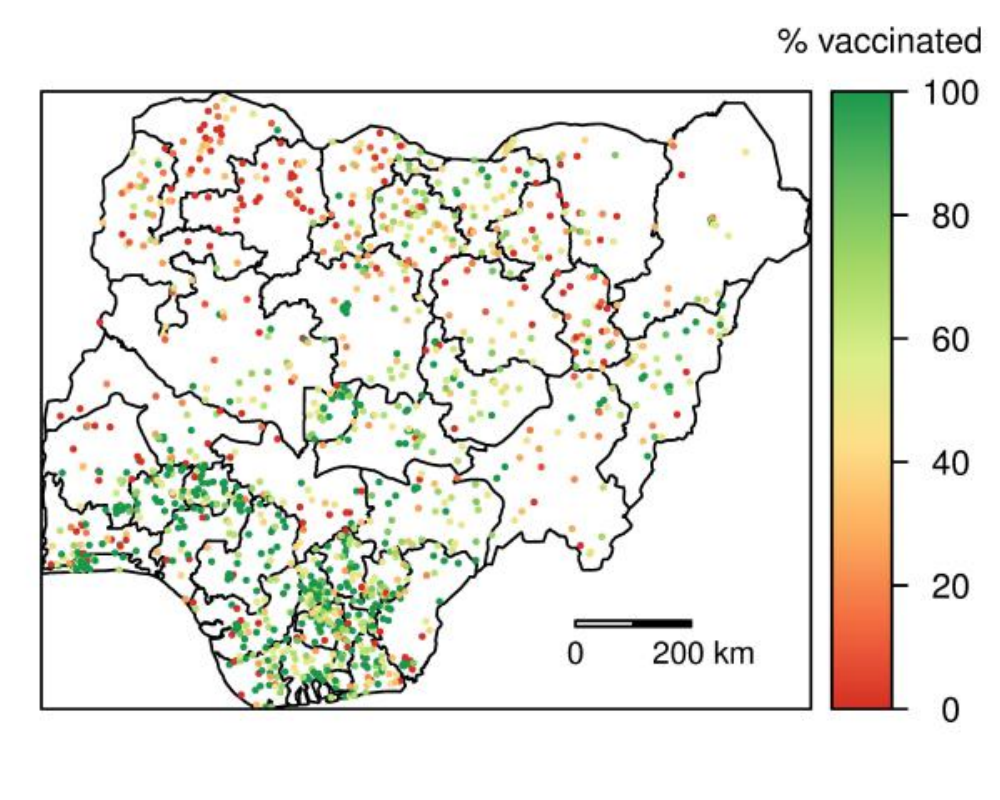}
	\caption{Spatial distribution of the observed MCV1 coverage among children aged 12--23 months as recorded at the 2018 NDHS cluster level. The cluster-level observed MCV1 coverage is calculated as the proportion of children sampled in a survey cluster who have had at least one dose of MCV at the time of interview, based on evidence from either vaccination cards or caregiver recall.}
	\label{figure1}
\end{figure}

The 2018 NDHS used a stratified, two-stage cluster design and a sampling frame derived from the Nigeria census conducted in 2006. Stratification was achieved by separating each of the 36 states and the Federal Capital Territory (i.e., the 37 administrative-1 areas, and hereafter referred to as Nigeria's 37 states) into urban and rural areas. Samples were selected independently in each stratum via a two-stage process: first, a pre-specified number of primary sampling units (PSUs), referred to as survey clusters, were selected from the sampling frame of census enumeration areas (EAs) with probability proportional to size; then, a fixed number of 30 households in every cluster were selected through equal probability systematic sampling. A total of 1389 survey clusters were selected to provide results representative at the national level as well as the state level. In this paper, we focus on the coverage of MCV1 among children aged 12--23 months based on evidence from either vaccination cards or caregiver recall. We remove the clusters that had no GPS location information or no eligible child samples and analyze data collected from the 5886 children aged 12--23 months in the remaining 1301 survey clusters. Figure \ref{figure1} shows the spatial distribution of the observed MCV1 coverage among children aged 12--23 months as recorded at the cluster level. In general, the survey clusters in northern Nigeria tend to have lower observed MCV1 coverage than those in the south, and clusters that are closer to each other tend to have similar observed coverage -- a sign of spatial correlation. Additional details of the survey data and exploratory analysis results can be found in the supplementary materials.

\begin{center}
    [Figure \ref{figure1} about here]
\end{center}


\section{Acknowledging the survey design in modeling} \label{sec_mod}

The stratified multi-stage cluster design adopted by the 2018 NDHS is ubiquitous among major household surveys that measure vaccination coverage. When fitting continuous spatial models to data from such surveys, it is essential to account for two key characteristics of the design: the stratification and clustering of the samples. A simulation study conducted by Paige \etal~\cite{paige2019design} shows that explicitly accounting for survey stratification by including urbanicity as a covariate in continuous spatial regression models improves predictions. If the response is associated with urbanicity then the improvement can be considerable. If any strong urban/rural association with the outcome is not properly modeled, large bias can result. WorldPop and IHME do not explicitly adjust for stratification in their models \cite{utazi2018high, utazi2018spatial, utazi2019mapping, mosser2019mapping, utazi2020geospatial}, but they do include extensive covariates which may, to some extent, implicitly adjust for the urban/rural stratification. 

Acknowledging the clustering of survey samples is also challenging. WorldPop routinely ignore clustering in their models \cite{utazi2018high, utazi2018spatial,utazi2019mapping}. IHME include an independent \textit{nugget} error term in the linear predictor of cluster-level coverage estimate to capture the non-spatial excess variation in survey outcomes \cite{mosser2019mapping}. However, this approach has been used without explicit consideration of the mechanisms by which excess variation at the cluster level manifests itself. Traditionally, in the context of a continuous outcome, an estimated \textit{nugget} effect has been attributed to measurement error or small-scale spatial variation \cite{diggle2019model, chiles2009geostatistics}. In the context of a vaccination coverage survey where Bernoulli sampling is carried out, measurement error would correspond to misclassification (willfully or by accident) of the binary outcomes. This undoubtedly occurs, but explicitly modeling the misclassification probabilities is difficult without gold standard data. In this paper, we focus on two other plausible forms of excess variation: (i) \textbf{within-cluster} variation that induces \textbf{overdispersion} at the cluster level, and (ii) \textbf{between-cluster} variation that represents \textbf{true signal} at the cluster level. The former is often referred to as excess-binomial variation, and the latter are ``shocks" in the signal due to specific conditions at the cluster. One needs to carefully consider which form is adopted before choosing an estimation procedure that gives appropriate inference. 

To illustrate and compare the various approaches of accounting for survey design in modeling, we consider variants of the following base model in which neither stratification nor clustering is explicitly acknowledged: 
\begin{align}
& Y_{ic} | p_{ic} \sim \text{Binomial} \left( n_{ic}, p_{ic} \right)\\
& \logit ( p_{ic} ) = \alpha + \bm{\beta}^\top \bm{x}_{ic} + S(\bm{s}_{ic}) \label{eqn2}
\end{align}
Here, $Y_{ic}$ is the random variable representing the number of vaccinated children out of $n_{ic}$ who are sampled in cluster $c$ of state $i$, $p_{ic}$ is the vaccination coverage parameter, and $\bm{x}_{ic}$ is the vector of covariates associated with cluster $c$ in state $i$. We let $\bm{s}_{ic}$ be the location associated with cluster $c$ in state $i$ and $S(\bm{s}_{ic})$ be a spatial random effect that follows a GP: $S(\cdot) \sim \text{GP}\left(0, \bm{\Sigma}_S \right)$ with $\bm{\Sigma}_S = \left[ \sigma_{S}^2, \rho \right]$, where $\sigma_{S}^2$ is the marginal spatial variance and $\rho$ is the spatial range (i.e., a distance at which the spatial correlation becomes negligible). Note that the GP we use is the solution to a stochastic partial differential equation (SPDE) which is approximated by a particular Gaussian Markov Random Field (GMRF) defined on a fine triangular mesh \cite{lindgren2011explicit}. This binomial model includes \textbf{no nugget} at the cluster level beyond the spatial field, hence we label this the \textit{Binomial NN} model. 

To obtain model-based vaccination coverage estimate for cluster $c$ in state $i$, we can construct an approximation to its marginal posterior distribution by drawing posterior samples: 
\begin{align}
p_{ic}^{(m)} =  \expit \left( \alpha^{(m)} + \bm{\beta}^{\top (m)} \bm{x}_{ic} + S(\bm{s}_{ic})^{(m)} \right), \label{eqn3}
\end{align}
where the superscript $(m)$ denotes the $m$th posterior sample of the respective parameter. Recall that a survey cluster is a census EA that has been sampled for the survey. Therefore, administrative-area-level coverage estimates can be obtained by aggregating the cluster-level estimates of all EAs in the area weighted by the population proportion in the EA. For example, the model-based vaccination coverage estimate for state $i$ can be calculated using the posterior samples
\begin{align*}
p_{i}^{(m)} =  \sum_{c=1}^{C_i} p_{ic}^{(m)} \times q_{ic}, 
\end{align*}
where $C_i$ is the total number of EAs in state $i$ and $q_{ic}$ is the proportion of the 12--23m population of state $i$ that is in EA/cluster $c$. Almost always, however, the complete sampling frame of EAs is not available. In this case, we can approximate the state-level coverage by aggregating cluster-level estimates over an approximated gridded EA map:
\begin{align}
p_{i}^{(m)} \approx  \sum_{g=1}^{G_i} p_{ig}^{(m)} \times q_{ig}, \label{eqn4}
\end{align}
where $g$ indexes the gridded EA and $G_i = C_i$ is the total number of EAs in state $i$. The approximated EA map can be created using a gridded population density map and summary tables that are routinely available from survey reports. Details are provided in the supplementary materials.

To acknowledge the urban/rural stratification, we can extend (\ref{eqn2}) to include urbanicity as a covariate in the model: 
\begin{align}
\logit ( p_{ic} ) = \alpha + \bm{\beta}^\top \bm{x}_{ic} + \gamma I\left( \bm{s}_{ic} \in \text{Urban} \right)+ S(\bm{s}_{ic}).
\end{align}
Inference for cluster-level coverage $p_{ic}$ can be obtained using posterior samples of contributing parameters in an analogous way to equation (\ref{eqn3}), and the administrative-area-level estimates can be approximated using the aggregation procedure summarized in equation (\ref{eqn4}).

Now we focus on accounting for clustering and consider two mechanisms by which cluster-level non-spatial variation arises. 
\begin{enumerate} 

\item \textbf{Within-Cluster Variation}. Suppose within a super-population in a cluster there are groups with their own distinct vaccination coverage. If we carried out repeated sampling at this cluster, a different group (or mixtures of groups) would be sampled each time and we would ``see" a different coverage beyond sampling variation. This phenomenon has been extensively studied in the statistics literature \cite{mccullagh1989generalized}. In this case, we have within-cluster variation that induces \textbf{overdispersion (OD)} at the cluster level. What we are really interested in is the overall coverage we would see if we could sample everyone in the cluster. To capture this kind of variation, we can assume that each group's coverage is drawn from a distribution. We describe two possible models that achieve this purpose. First, we can assume that each group's coverage follows a beta distribution. This results in a Beta-Binomial likelihood for the data:
\begin{align}
& Y_{ic} | \mu_{ic}, d \sim \text{Beta-Binomial} \left( n_{ic}, \mu_{ic}, d \right) \\
& \logit(\mu_{ic}) = \alpha + \bm{\beta}^\top \bm{x}_{ic} + \gamma I\left( \bm{s}_{ic} \in \text{Urban} \right)+ S(\bm{s}_{ic}). \label{eqn7}
\end{align}
Specifically, we can think of this model as a result of:
\begin{align*}
&Y_{ic} | p_{ic} \sim \text{Binomial} \left( n_{ic}, p_{ic} \right) \\
&p_{ic} | \mu_{ic}, d \sim \text{Beta}\left( \mu_{ic}, d \right),
\end{align*}
where the beta distribution is parameterized as 
\begin{align*}
E\left[ p_{ic} | \mu_{ic}, d \right] &= \mu_{ic} \\
var\left[ p_{ic} | \mu_{ic}, d \right] &= \frac{\mu_{ic} (1- \mu_{ic})} {d + 1}.
\end{align*}
Marginally, the data follows a Beta-Binomial distribution with 
\begin{align*}
E\left[ Y_{ic} | \mu_{ic}, d \right] &= n_{ic} \mu_{ic}\\
var\left[ Y_{ic} | \mu_{ic}, d \right] &= n_{ic} \mu_{ic} (1- \mu_{ic}) \times \frac{n_{ic} + d}{1+d}.
\end{align*} 
The parameter $d > 0$ characterizes the degree of overdispersion, with higher values of $d$ corresponding to less overdispersion\footnote{The limiting case of a Beta-Binomial distribution at $d = \infty$ is a binomial distribution.}.  The target of inference is $E\left[ p_{ic} | \mu_{ic}, d \right] = \mu_{ic}$, which corresponds to the hypothetical coverage we would see if we could sample everyone in the cluster -- it can be estimated using posterior samples of contributing parameters in (\ref{eqn7}). We label this model the \textit{Beta-Binomial OD} model. 

An alternative is to assume that each group's coverage follows a normal distribution on the logit scale (i.e., a logit-normal distribution). This results in a Lono-Binomial\footnote{This is a made-up name for the compound distribution where one can think of the $p$ parameter in the binomial distribution as being randomly drawn from a logit-normal distribution. We want to emphasize the parallel between this distribution and the Beta-Binomial distribution. Details are provided in the supplementary materials.} likelihood for the data: 
\begin{align*}
& Y_{ic} | \eta_{ic}, \sigma^2_{\delta} \sim \text{Lono-Binomial} \left( n_{ic}, \eta_{ic}, \sigma^2_{\delta} \right) \\
& \eta_{ic} = \alpha + \bm{\beta}^\top \bm{x}_{ic} + \gamma I\left( \bm{s}_{ic} \in \text{Urban} \right)+ S(\bm{s}_{ic}).
\end{align*}
Specifically, we can think of this model as a result of:
\begin{align*}
&Y_{ic} | q_{ic} \sim \text{Binomial} \left( n_{ic}, q_{ic} \right) \\
&\logit\left( q_{ic} \right)| \eta_{ic}, \sigma^2_{\delta} =\eta_{ic} + \delta_{ic}\\
&\delta_{ic} \sim_{iid} \text{Normal}\left( 0, \sigma^2_{\delta} \right)
\end{align*}
The parameter $\sigma^2_{\delta} > 0$ characterizes the degree of overdispersion, with higher $\sigma^2_{\delta}$ value corresponding to more overdispersion\footnote{The limiting case of a Lono-Binomial distribution at $\sigma^2_{\delta} = 0$ is a binomial distribution.}.  The target of inference in this case is $E \left[ q_{ic}| \eta_{ic}, \sigma^2_{\delta} \right]$, which, by the law of the unconscious statistician, equals
\begin{align*}
p_{ic} = E \left[ q_{ic}| \eta_{ic}, \sigma^2_{\delta} \right] &= \int_{\delta} \expit \left( \eta_{ic} + \delta \right) \pi \left( \delta | \sigma^2_{\delta} \right) d\delta \\
&\approx \expit \left( \frac{\eta_{ic}}{\sqrt{1 + h^2\sigma^2_{\delta}}} \right) \\ 
& = \expit \left( \frac{\alpha + \bm{\beta}^\top \bm{x}_{ic} + \gamma I\left( \bm{s}_{ic} \in \text{Urban} \right)+ S(\bm{s}_{ic})}{\sqrt{1 + h^2\sigma^2_{\delta}}} \right), \numberthis \label{eqn8}
\end{align*}
where $h = \frac{16\sqrt(3)}{15\pi}$. For a deviation of this approximation, see Section 9.13.1 of Wakefield (2013) \cite{wakefield2013bayesian}. Inference for this target can be obtained using posterior samples of contributing parameters in (\ref{eqn8}). We label this model the \textit{Lono-Binomial OD} model. 

\item \textbf{Between-Cluster Variation.} The cluster-level excess variation may also be \textbf{true signal (TS)} that represents differences in cluster coverage means (beyond the spatial field). In this case, if we carried out repeated sampling at this cluster, we would ``see" different groups having the same mean coverage. We can model this kind of variation using an independent normal term in the linear predictor of cluster-level vaccination coverage:
\begin{align*}
& Y_{ic} | p_{ic} \sim \text{Binomial} \left( n_{ic}, p_{ic} \right) \\
& \logit (p_{ic}) = \alpha + \bm{\beta}^\top \bm{x}_{ic} + \gamma I\left( \bm{s}_{ic}  \in \text{Urban} \right) + S(\bm{s}_{ic}) + \epsilon_{ic} \\
& \epsilon_{ic} \sim \text{Normal} \left(0, \sigma^2_{\epsilon} \right) 
\end{align*}
We emphasize that $\epsilon_{ic}$ is capturing between-cluster differences in coverage, not the within-cluster variability that induces overdispersion in the \textit{Lono-Binomial OD} model. Hence, the target of inference is 
\begin{align}
p_{ic} = \expit\left( \alpha + \bm{\beta}^\top \bm{x}_{ic} + \gamma I\left( \bm{s}_{ic}  \in \text{Urban} \right) + S(\bm{s}_{ic}) + \epsilon_{ic} \right), \label{eqn9}
\end{align}
which is different from (\ref{eqn8}). Posterior samples of contributing parameters can be used to estimate (\ref{eqn9}), with $\epsilon_{ic}^{(m)}$ randomly sampled from a $\text{N} \left(0, \sigma^{2(m)}_{\epsilon} \right)$ distribution. We label this model the \textit{Binomial TS} model. 
\end{enumerate}

In reality, the cluster-level non-spatial variation is likely to be the result of a mixture of the within-cluster overdispersion and between-cluster true signal, but most household surveys only obtain a single sample of households within each selected cluster, so the survey data does not contain enough information to identify the source of the cluster-level variation. Therefore, map producers need to be extremely careful about their assumptions regarding cluster-level extra variation when choosing the target of inference for vaccination coverage estimation. 

We apply the aforementioned four classes of models, namely \textit{Binomial NN}, \textit{Beta-Binomial OD}, \textit{Lono-Binomial OD} and \textit{Binomial TS}, to analyze data from the 2018 NDHS. We fit two models within each class: one with urbanicity as a covariate to account for survey stratification and the other without. In addition, all models include the same set of covariates: poverty, aridity, log-transformed night-time lights, log-transformed travel time and enhanced vegetation index (EVI). We use the widely applicable information criterion (WAIC) \cite{watanabe2010asymptotic} to evaluate the predictive power of the models, and conduct a 37-fold cross-validation exercise, holding out data from one whole state each time, to assess the performance of out-of-sample predictions. We also calculate the bias, mean absolute error (MAE) and root mean squared error (RMSE) for each model, all of which are described in the supplementary materials. 

All models were fitted using the INLA approach \cite{rue2009approximate} implemented in the \texttt{INLA} package in \texttt{R}. Table \ref{table1} reports the estimates of selected parameters and the model validation results for each model. Based on the results, the odds of being vaccinated against measles is estimated to be 35\% to 42\% higher in an urban area than in a rural area. The 95\% credible intervals (CIs) of the corresponding coefficients are always strictly greater than 0 (i.e., odds ratio greater than 1), indicating a strong association between urbanicity and MCV1 coverage after accounting for the other covariates. Across all model classes, adding the urban/rural strata variable to account for survey design always improves predictive performance in terms of WAIC, bias, MAE and RMSE. This result is consistent with the findings of Paige \etal~\cite{paige2019design}. The estimated coefficients for the other covariates are reported in the supplementary materials and the results are fairly consistent across all models. 

\begin{center}
    [Table \ref{table1} about here]
\end{center}

We now compare the \textit{Binomial NN} models, which assume there is no cluster-level excess variation beyond the spatial field, to the other three classes of models, which explicitly account for non-spatial excess variation under various assumptions. The estimated spatial fields from the \textit{Binomial NN} models have shorter range and smaller marginal spatial variance, which characterize a ``bumpy" surface  with a lot of local residual spatial correlation (the spatial range $\approx 78$ km). In contrast, the estimated spatial fields from the other models tend to be ``smoother" with spatial correlation over longer distances (spatial range $\approx 322$ km). This contrast hints at the presence of considerable small-scale variation that can be better captured by the \textit{Beta-Binomial OD}, \textit{Lono-Binomial OD} and \textit{Binomial TS} models. The estimates for the overdispersion/true signal parameters in these models indicate significant cluster-level variation beyond the spatial field. In addition, their predictive performances measured by WAIC, Bias, MAE and RMSE are noticeably better than that of the \textit{Binomial NN} models (see Table \ref{table1}). 
In summary, this analysis illustrates the advantages of accounting for the survey design, especially the survey stratification and clustering, when fitting geostatistical models to estimate vaccination coverage. Once a satisfactory model is selected, it is important to present the model-based estimates with care. This will be the focus of the next section. 


\section{Presenting model-based estimates and their uncertainties} \label{sec_pres}

\begin{figure}[H]
	\centering
	\begin{adjustwidth}{-3cm}{-3cm}
	\includegraphics[width=1.5\textwidth]{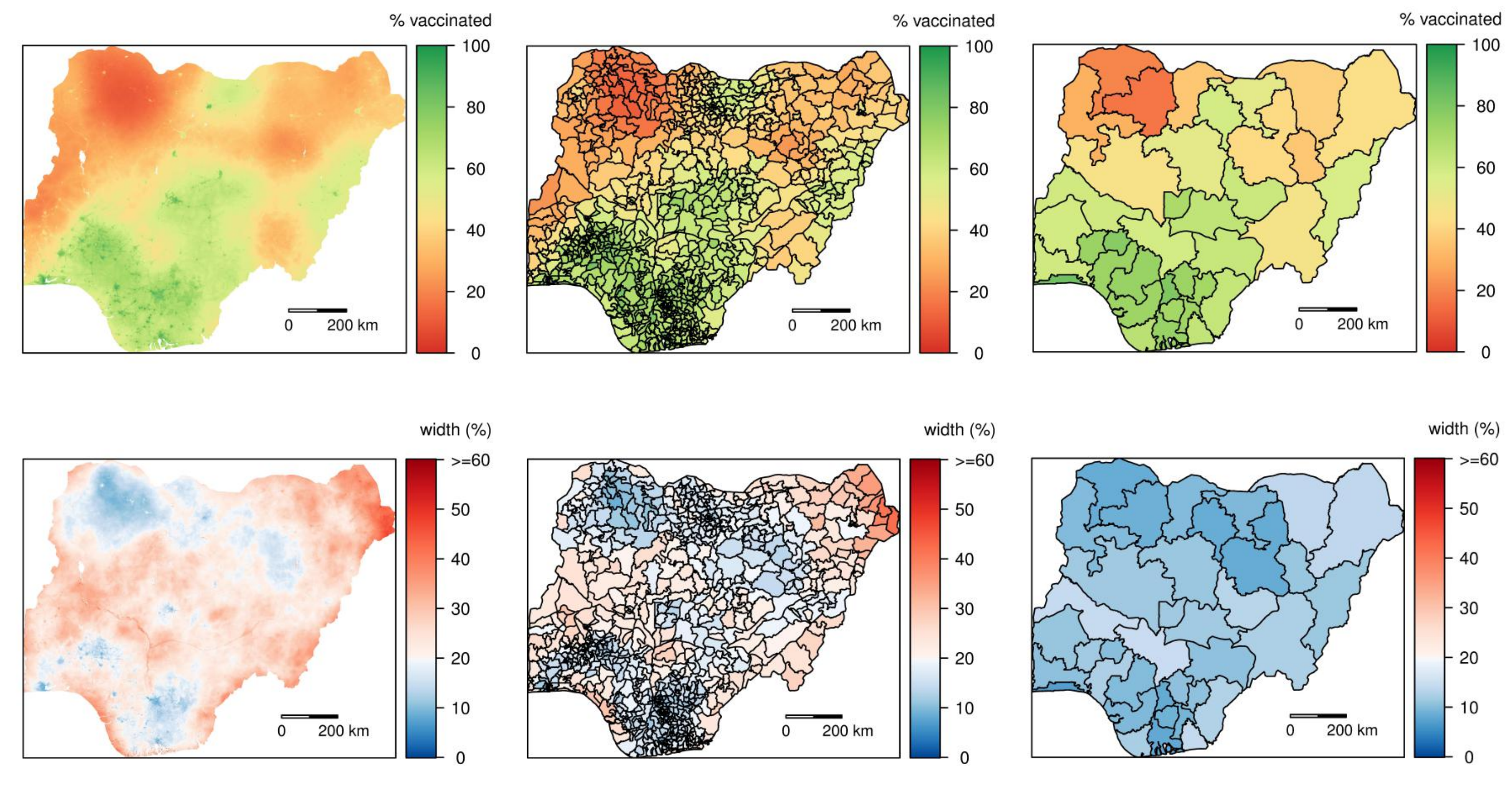}
	\end{adjustwidth}
	\caption{Maps of the posterior medians (top row) and the widths of 90\% credible intervals (botton row) for the estimated MCV1 coverage at the $1\times 1$ km pixel (left), LGA (middle) and state (right) levels, based on the  \textit{Lono-Binomial OD} model that includes the urban/rural strata variable.}
	\label{figure2}
\end{figure}

As mentioned in Section \ref{sec_intro}, pixel-level vaccination coverage estimates are routinely produced as output of geostatistical models, and they are often aggregated by population density to form administrative-area-level coverage estimates. Figure \ref{figure2} shows the maps of the posterior medians and the widths of 90\% credible intervals (CIs) for the estimated MCV1 coverage at the $1\times 1$ km pixel, local government area (LGA, which is administrative-2 area) and state levels based on the \textit{Lono-Binomial OD} model that includes the urban/rural strata variable. The plots derived from other models can be found in the supplementary materials. These maps show a consistent trend: coverage estimates at a finer spatial resolution tend to have larger associated uncertainty --- and hence poorer precision. In particular, the pixel maps are often associated with huge uncertainties --- a consequence of the sparsity of the survey data, since the 2018 NDHS was only powered to be representative at the state level \cite{dhs2018}. Although geostatistical models can usually produce smoothed estimates that are slightly biased but more precise than the direct estimates (Figure S1 in the supplementary materials), the $1\times 1$ km pixel is too fine a spatial resolution for reliable coverage estimation using the typical number of clusters in household surveys. Therefore, when presenting vaccination coverage estimates on a map, one needs to carefully consider the trade-off between the geographical scale and precision of estimates, and choose an appropriate spatial resolution such that the resultant map have reasonably high precision for the estimates to be statistically reliable. In addition, the uncertainty associated with the point estimates should always be appropriately presented. Instead of labeling uncertainty maps with ambiguous terms such as ``low" and ``high" \cite{utazi2018high, utazi2019mapping, utazi2020geospatial}, quantitative scales should be used to clearly show how reliable the coverage estimates are. Examples of commonly used quantitative uncertainty measures include posterior CI width and standard deviation. Another sensible choice is the coefficient of variation (CV) of the posterior distribution. We provide more details regarding the use of CV in the supplementary materials.

\begin{center}
    [Figure \ref{figure2} about here]
\end{center}

\begin{figure}[H]
	\centering
	\begin{adjustwidth}{-2.5cm}{-2.5cm}
	 \includegraphics[width=1.3\textwidth]{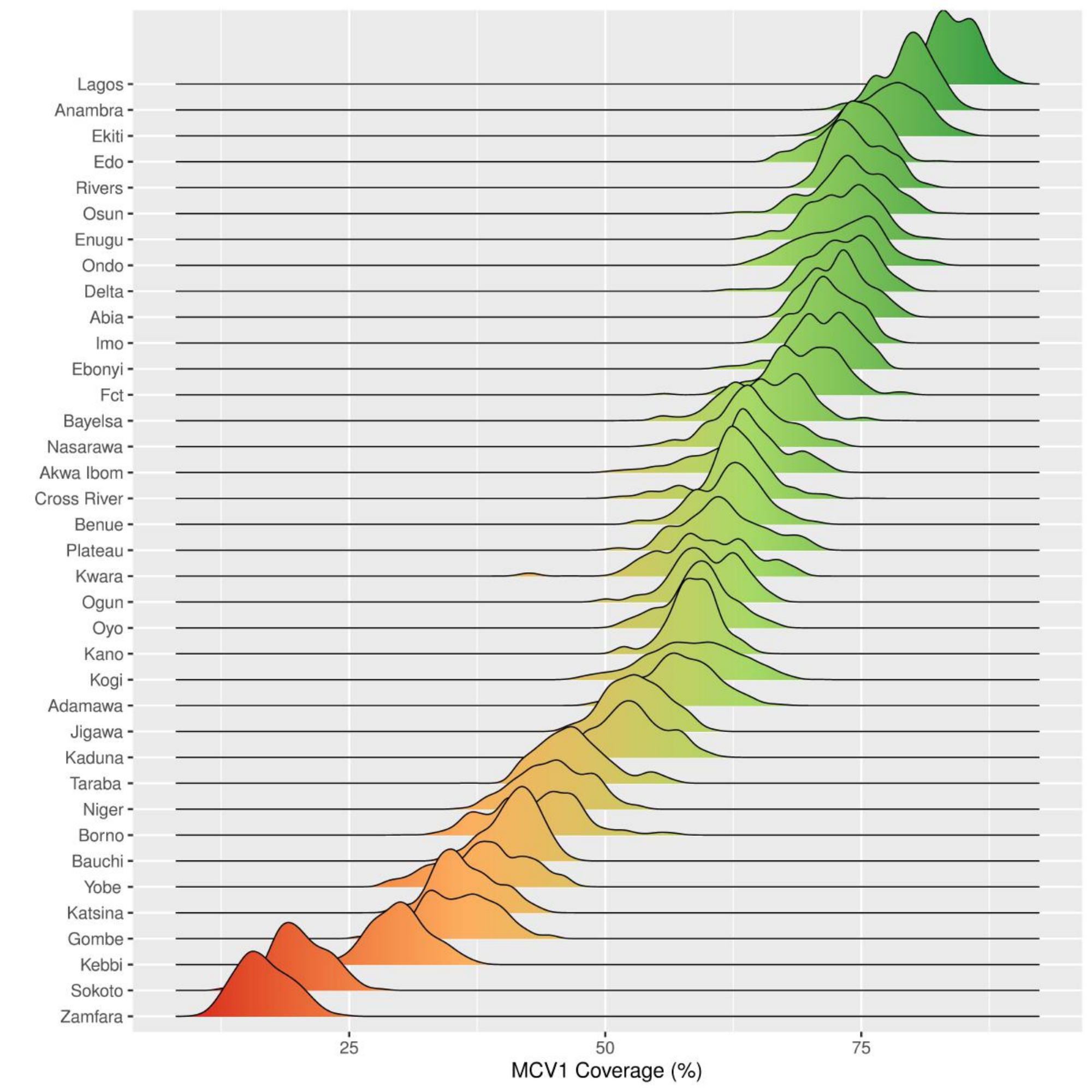}
	\end{adjustwidth}
	\caption{Ridgeline plots of the posterior distributions of the MCV1 coverage estimates for Nigeria's 37 states, ordered by posterior median, based on 1000 posterior samples from the  \textit{Lono-Binomial OD} model that includes the urban/rural strata variable. }
	\label{figure3}
\end{figure}

An important aim of vaccination coverage estimation is to rank a set of areas to identify the places that need the most improvement. Often, a single summary statistic, such as the posterior mean or median of the vaccination coverage, is used for ranking areas with little consideration of the associated uncertainties. To better visualize the uncertainty associated with the ranking, we recommend two presentation methods. The first is to use ridgeline plots to show the posterior distribution of coverage estimate for each area. For example, Figure \ref{figure3} shows the ridgeline plots of the posterior distributions of the MCV1 coverage estimates for Nigeria's 37 states, ordered by posterior median, based on 1000 posterior samples from the  \textit{Lono-Binomial OD} model that includes the urban/rural strata variable. When comparing MCV1 coverage estimates across different areas, one can clearly see how much the areas' posterior distributions overlap with each other on the ridgeline plots, and hence be informed of the uncertainty associated with the relative ranking of the areas.

\begin{center}
    [Figure \ref{figure3} about here]
\end{center}

\begin{figure}[H]
	\centering
	\begin{adjustwidth}{-0.5cm}{-0.5cm}
	 \includegraphics[width=1.1\textwidth]{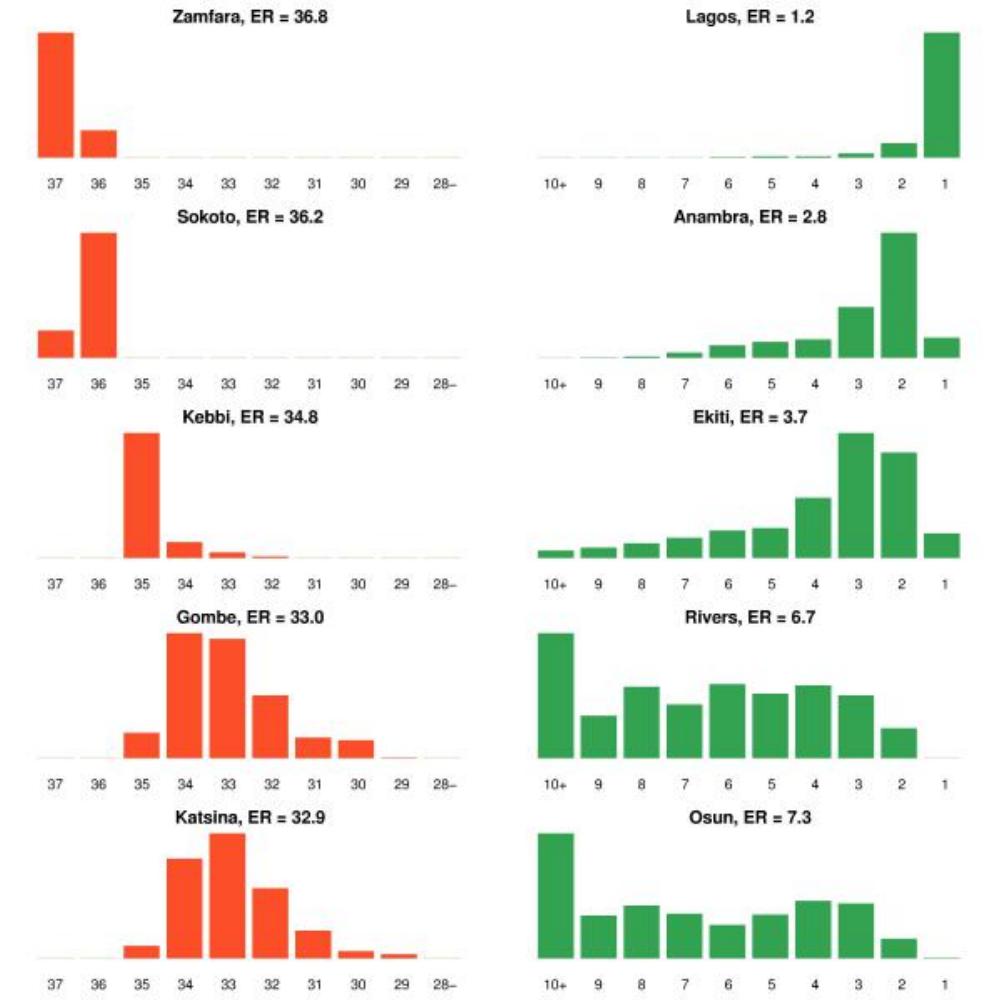}
	\end{adjustwidth}
	\caption{Histograms of the posterior ranking distributions of the 5 states with the lowest (left column) and 5 with the highest (right column) expected ranks (ERs), based on 1000 posterior samples of the MCV1 coverage estimates of Nigeria's 37 states from the  \textit{Lono-Binomial OD} model that includes the urban/rural strata variable. }
	\label{figure4}
\end{figure}

Alternatively, one can use histograms of the posterior ranking distributions to identify areas with the highest or lowest vaccination coverage. For a given set of areas, each Monte Carlo draw of posterior coverage estimates gives a ranking of the areas. One can therefore obtain a posterior ranking distribution for each area by repeatedly drawing and ranking posterior samples, and the expected rank (ER) for each area can be estimated by taking the average of its ranking samples. For example, Figure \ref{figure4} shows the histograms of the posterior ranking distributions of the 5 states with the lowest and 5 with the highest ERs based on 1000 posterior Monte Carlo samples of the state-level MCV1 coverage estimates. Zamfara, Sokoto and Kebbi seem to have the highest posterior probabilities of being ranked as the lowest three states. As for Gombe and Katsina, the similarity in their ranking distributions reveals high uncertainty in their relative rankings. On the other hand, the state of Lagos is likely to have the highest MCV1 coverage, followed by the states of Anambra and Ekiti. However, there is considerable uncertainty in the ranking of the next two states: Rivers and Osun, as is evident in the wide spread of their ranking distributions. In summary, both the ridgeline plots of posterior distributions and histograms of posterior ranking distributions are useful tools for identifying areas with the highest or lowest vaccination coverage. They are not only straightforward to calculate, but also effective in visualising the uncertainty associated with the ranking of the areas. 

\begin{center}
    [Figure \ref{figure4} about here]
\end{center}

\begin{figure}[H]  
	\centering
	\begin{adjustwidth}{-3cm}{-3cm}
	 \includegraphics[width=1.5\textwidth]{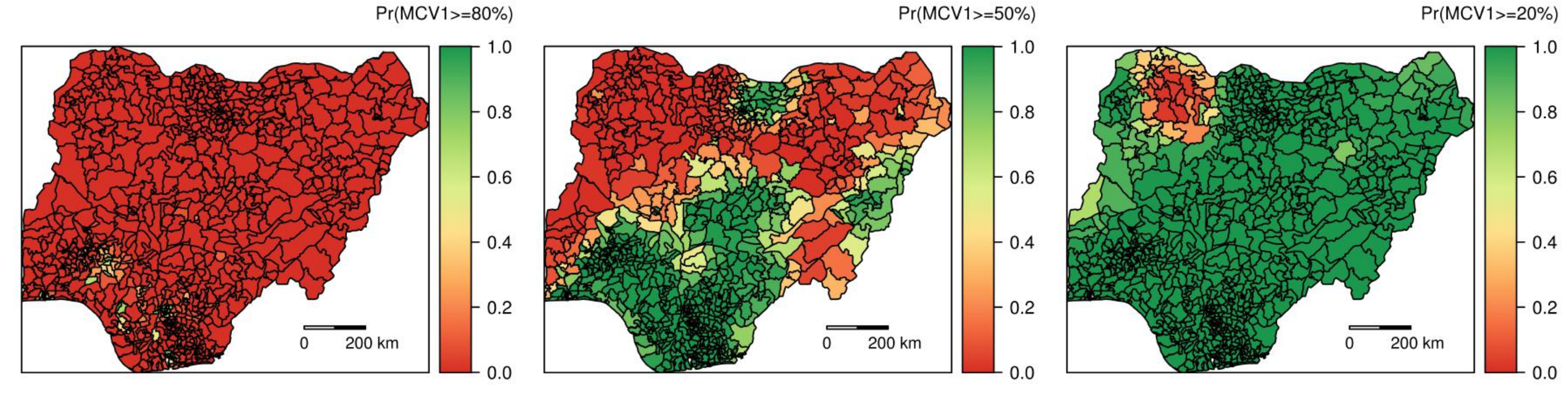}
	\end{adjustwidth}
	\caption{Maps of the posterior probabilities of LGA-level MCV1 coverage estimates exceeding the 80\%, 50\% and 20\% thresholds respectively, based on the  \textit{Lono-Binomial OD} model that includes the urban/rural strata variable.}
	\label{figure5}
\end{figure}

Another common objective of vaccination coverage estimation is to identify areas with vaccination coverage higher or lower than a certain threshold. In this case, it is useful to examine maps of posterior exceedance probabilities. For example, Figure \ref{figure5} shows the maps of the marginal posterior probabilities of LGA-level MCV1 coverage exceeding the 80\%, 50\% and 20\% thresholds respectively. In particular, the exceedance probability map corresponding to the 80\% threshold is especially relevant in evaluating to what extent Nigeria has attained the Global Vaccine Action Plan (GVAP) target of reaching 80\% coverage with all vaccines in all districts by 2020 \cite{GVAP}. The map highlights that substantial efforts are needed in most places to meet the target. Fewer than 7 percent of the LGAs (i.e., 53 out of 774 LGAs) have 50\% or greater posterior probability of being at the target of 80\% MCV1 coverage. In addition, the exceedance probability map corresponding to the 20\% threshold identifies 28 LGAs in northwest Nigeria that have 50\% or lower posterior probability of reaching an MCV1 coverage level of 20\%. This demonstrates how exceedance probability maps can be used to identify areas with low coverage estimates while keeping users informed of the uncertainties associated with the results. 

\begin{center}
    [Figure \ref{figure5} about here]
\end{center}

In addition to visualizing the uncertainties associated with coverage estimates, one should also avoid showing maps with low overall precision. We propose a novel approach to coverage map presentation that allows comparison and control of the overall map uncertainty level. Our method is inspired by Bayesian decision theory and is applicable to presenting coverage estimates at any spatial scale.

The basic idea is that we move away from the usual continuous color scales; instead, we use a discrete set of colors to represent a partition of $[0\%, 100\%]$, the range of vaccination coverage. The partition can be defined based on some pre-specified threshold(s) of interest. For example, we can use the thresholds 20\%, 50\% and 80\% to create four intervals: $[0\%, 20\%)$, $[20\%, 50\%)$, $[50\%, 80\%)$ and $[80\%, 100\%]$, that represent the \textit{extremely low}, \textit{low}, \textit{medium} and \textit{high} coverage intervals respectively. Next, we examine the posterior distribution of the coverage estimate for each area and assign each area to the interval that contains the greatest posterior probability. We call this maximum the \textit{true classification probability} (TCP), so one minus the TCP is the probability of misclassificaition. The key idea here is that, as clearly shown in the ridgeline plots, for each color assigned to a pixel there is a posterior probability that this color is correctly classified but also a corresponding probability of incorrect classification. This step is inspired by Bayesian decision theory for univariate probabilities: the classification of each area follows a Bayes decision rule that minimizes a 0-1 loss. Finally, we calculate the average of the TCPs across all areas and call it the \textit{average true classification probability} (ATCP) of the map. The ATCP serves as a measure of the overall precision of a vaccination coverage map. A higher ATCP indicates a higher overall map precision, since on average we are more certain about the classification of the areas under the assumed model. 

\begin{figure}[H]  
	\centering
	\begin{adjustwidth}{-3cm}{-3cm}
	 \includegraphics[width=1.5\textwidth]{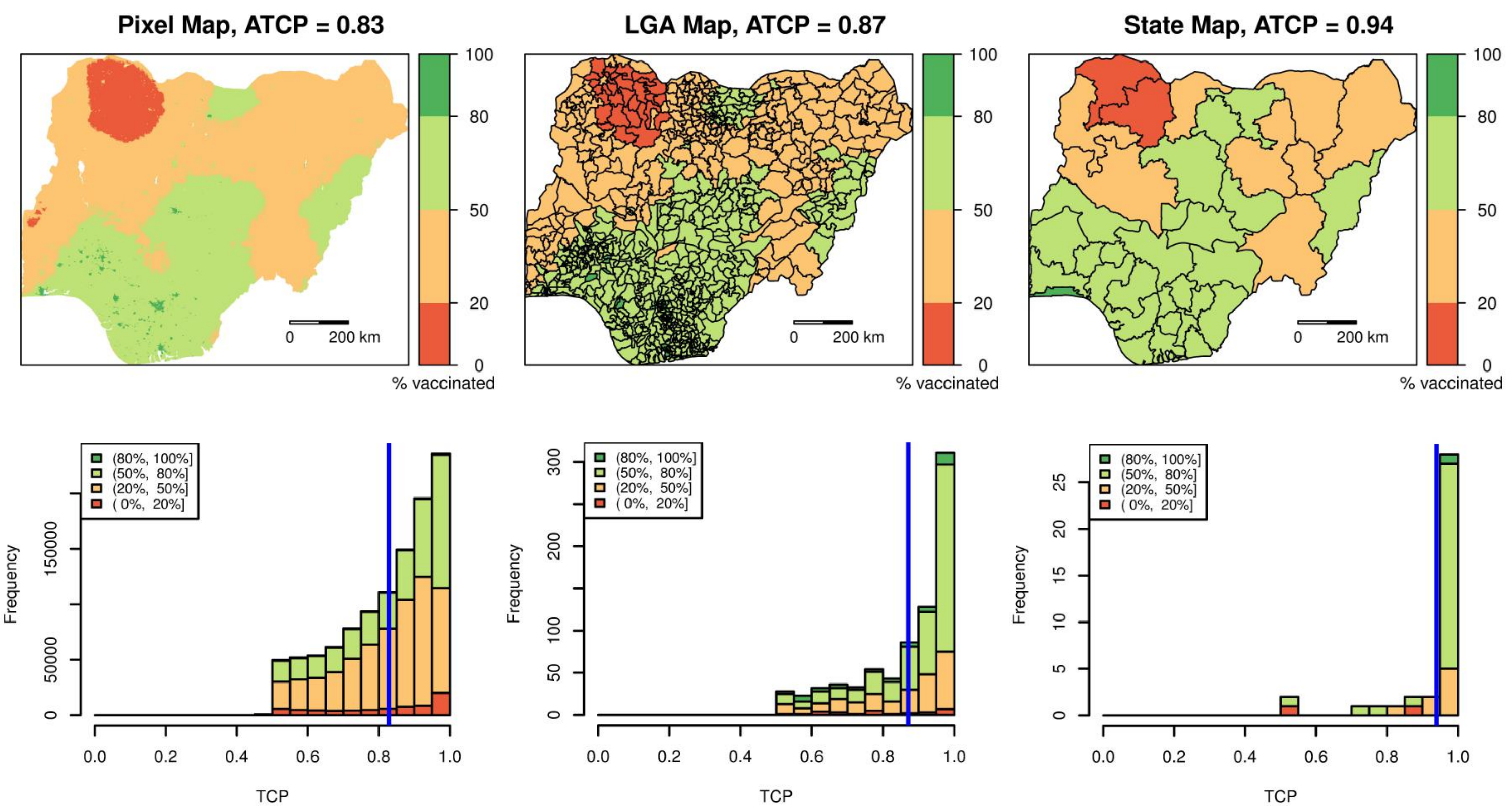}
	\end{adjustwidth}
	\caption{Top: Maps of MCV1 coverage estimates at the $1 \times 1$ km pixel, LGA and state levels using the discrete color scale composed of the \textit{extremely low} $[0\%, 20\%)$, \textit{low} $[20\%, 50\%)$, \textit{medium} $[50\%, 80\%)$ and \textit{high} $[80\%, 100\%]$ coverage intervals. Bottom: The corresponding histograms of the \textit{true classification probabilities} (TCPs) with the \textit{average true classification probability} (ATCP) highlighted by the blue vertical line. }
	\label{figure6}
\end{figure}

Using this approach, we present the MCV1 coverage estimates at the $1 \times 1$ km pixel, LGA and state levels based on the \textit{Lono-Binomial OD} model in Figure \ref{figure6}. The top row shows the maps with the pre-specified discrete color scale, and below each map is a histogram of the TCPs with the ATCP highlighted by a blue vertical line. We see that the state map has the highest ATCP (0.94), followed by the LGA map (ATCP = 0.87) and the pixel map (ATCP = 0.83). This reflects the decreasing precision associated with the state-, LGA- and pixel-level coverage estimates. 

\begin{center}
    [Figure \ref{figure6} about here]
\end{center}

\begin{figure}[H]  
	\centering
	\begin{adjustwidth}{-3cm}{-3cm}
	 \includegraphics[width=1.5\textwidth]{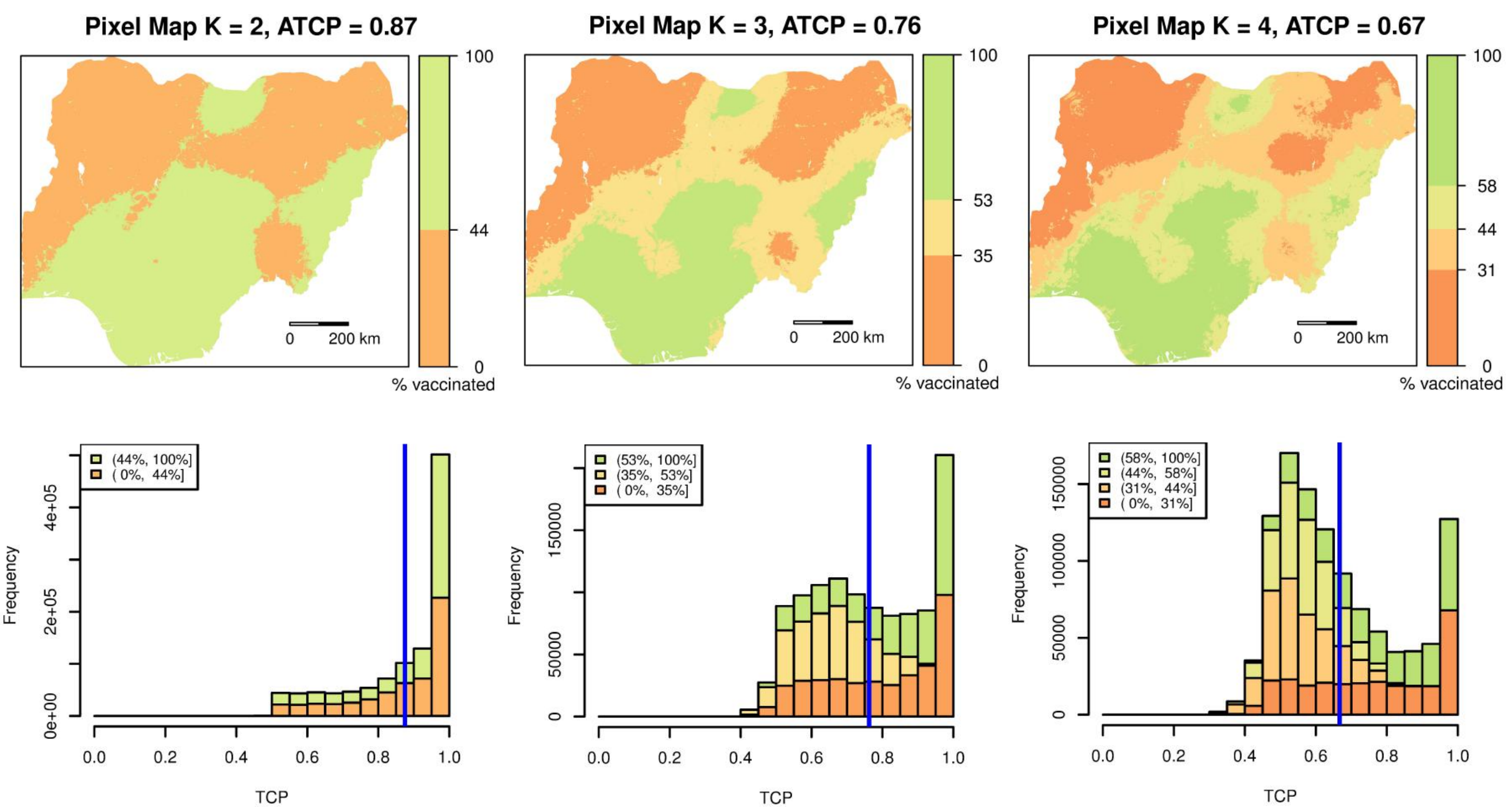} 
	\end{adjustwidth}
	\caption{Top: Maps of MCV1 coverage estimates at the $1 \times 1$ km pixel level using discrete color scales with $K = 2, 3 \mbox{ and } 4$ quantile intervals. For each $K$, the color scale is formed by creating the intervals $[L_0, L_1),\dots,[L_{K-1},L_K]$, where $L_0 = 0\%$, $L_K = 100\%$, and $L_k$ equals the $100 \times k/K$ quantile of the pooled posterior samples of the pixel-level coverage estimates based on the  \textit{Lono-Binomial OD} model that includes the urban/rural strata variable. Bottom: The corresponding histograms of the \textit{true classification probabilities} (TCPs) of the pixels with the \textit{average true classification probability} (ATCP) highlighted by the blue vertical line. }
	\label{figure7}
\end{figure}

If one does not have any specific threshold in mind to form a discrete color scale, one can consider using the quantiles of the posterior samples of the coverage estimates. We take the pixel-level MCV1 coverage estimates as an illustrate example: for a fixed number of levels $K$, one can form intervals $[L_0, L_1),\dots,[L_{K-1},L_K]$, where $L_0 = 0\%$, $L_K = 100\%$, and $L_k$ equals the $100 \times k/K$ quantile of the posterior samples pooled across all pixels. For example, for $K = 2$, we have $L_1 = 44\%$, the median of the pooled posterior samples, which forms two intervals: $[0\%, 44\%)$ and $[44\%, 100\%]$. Figure \ref{figure7} shows three pixel maps with discrete color scales formed by setting $K = $ 2, 3 and 4 respectively. Below each map is a histogram of the TCPs with the ATCP highlighted by a blue vertical line. We see that as the number of intervals $K$ increases, the ATCP becomes lower. This is expected: as we partition the range into more quantile intervals, the width of each interval would generally become narrower, and the TCPs would generally become smaller by definition. In fact, two extreme cases are possible when presenting model-based vaccination coverage estimates using this approach. At one extreme, a perfect ATCP can be achieved for showing a flat map where every pixel is assigned to the interval $[0\%,100\%]$. At the other extreme, we can assign each pixel to a unique color using a finely partitioned, almost-continuous color scale, but the ATCP would be essentially zero. The latter is the standard practice. 

\begin{center}
    [Figure \ref{figure7} about here]
\end{center}

To achieve a balance between the precision of a map and the granularity of its color scale, we can set a minimally acceptable level of ATCP and use the most granular color scale that satisfies this requirement. For example, if we set the minimally acceptable ATCP to be 0.70, then among the three pixel maps in Figure \ref{figure7}, the one corresponding to $K = 3$ would be the ``best" choice, since its ATCP (0.76) is above the 0.70 threshold and its color scale has more quantile intervals than the other eligible pixel map ($K=2$, ATCP = 0.87). The map corresponding to $K=4$ has an ATCP of 0.67, hence does not satisfy the minimally acceptable ATCP level of 0.70.

\begin{figure}[H]  
	\centering
	\begin{adjustwidth}{-0cm}{-0cm}
	 \includegraphics[width=1\textwidth]{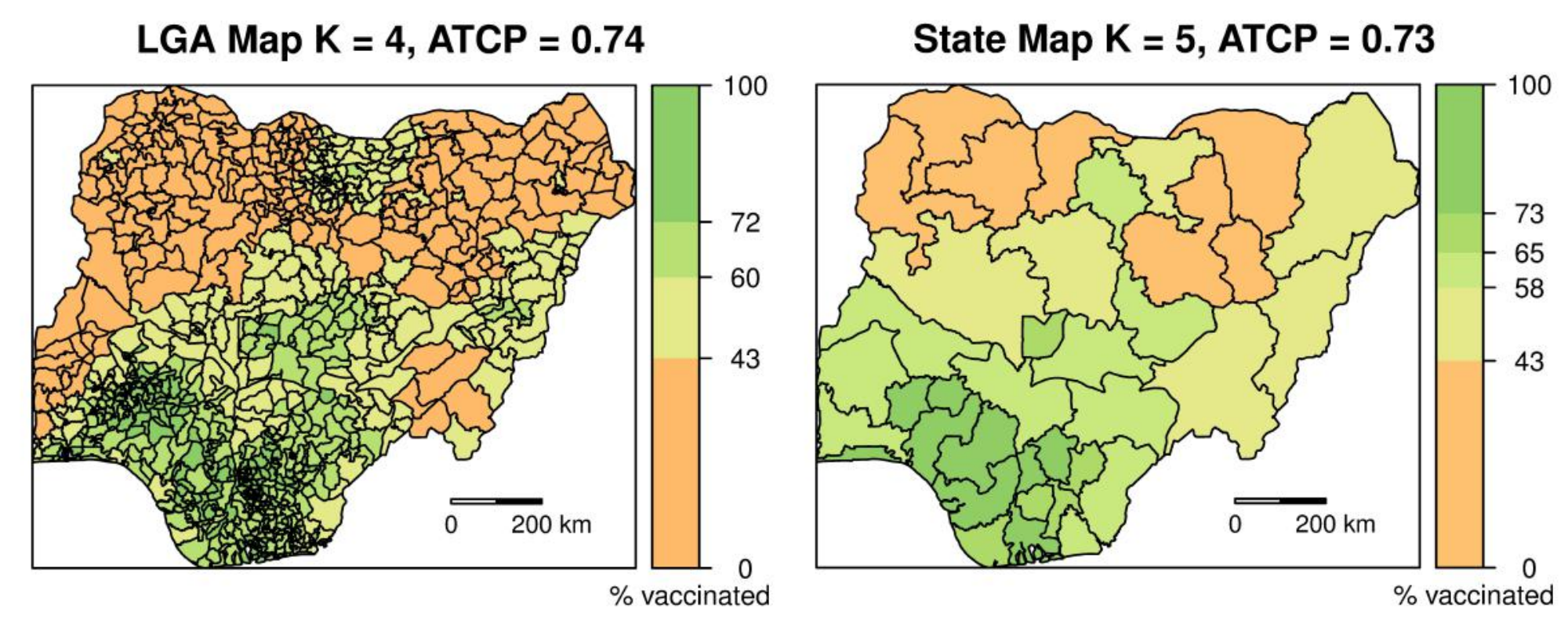}
	\end{adjustwidth}
	\caption{Left: Map of MCV1 coverage estimates at the LGA level using a discrete color scale with $K = 4$ quantile intervals. Right: Map of MCV1 coverage estimates at the state level using a discrete color scale with $K = 5$ quantile intervals. The color scales are formed by creating the intervals $[L_0, L_1),\dots,[L_{K-1},L_K]$, where $L_0 = 0\%$, $L_K = 100\%$, and $L_k$ equals the $100 \times k/K$ quantile of the pooled posterior samples of the coverage estimates at the LGA and state level respectively, based on the \textit{Lono-Binomial OD} model that includes the urban/rural strata variable.}
	\label{figure8}
\end{figure}

The same procedure can be applied to produce LGA- and state-level MCV1 coverage maps: if we set the same minimally acceptable ATCP at 0.70, we would have $K = 4$ and $5$ for the ``best" LGA and state maps respectively (Figure \ref{figure8}). This further illustrates the trade-off between the geographical scale and precision of estimates: given the same ATCP threshold, the state-level coverage estimates can be presented using a more granular color scale than the LGA- and pixel-level estimates, because the state-level estimates generally have higher precision. By setting a minimum threshold for ATCP, we can control the overall uncertainty level of a map, and hence avoid presenting maps with low overall precision.

\begin{center}
    [Figure \ref{figure8} about here]
\end{center}


\section{Discussion} \label{sec_disc}

In this paper, we have discussed a number of crucial choices with respect to the modeling and presentation of vaccination coverage estimates derived from household survey data. Using the MCV1 coverage among children aged 12--23 months in Nigeria as a motivating example, we fitted several Bayesian geostatistical models to the 2018 NDHS survey data and illustrated the importance of properly accounting for survey stratification and cluster-level non-spatial variation in survey outcome. In addition, we demonstrated the trade-off between the geographical scale and precision of estimates by showing the higher uncertainties associated with estimates at finer spatial resolutions. We also demonstrated several visualization methods for mapping and ranking that emphasize the probabilistic interpretation of results, including maps of posterior medians with CI widths, ridgeline plots of posterior distributions, histograms of posterior ranking distributions, and maps of posterior exceedance probabilities, Finally, a novel approach inspired by Bayesian decision theory is introduced to present vaccination coverage estimates using discrete color scales, allowing comparison and control of the overall map precision level.

Based on what we have discussed, we recommend the following guidelines for modeling and presenting vaccination coverage estimates using data from household surveys:
\begin{itemize}
    \item Survey stratification must be acknowledged in the model by including urbanicity as a covariate to avoid bias and improve predictive power.
    \item Sources of cluster-level non-spatial variation in survey outcomes must be carefully considered, and appropriate model and prediction scheme must be used to properly account for within-cluster variation that induces overdispersion or between-cluster variation that represents true signal.
    \item With most household surveys being designed to provide statistically reliable estimates at administrative-1 levels, it is critical to acknowledge the potentially high uncertainties associated with coverage estimates at fine spatial resolutions. Quantitative measures of uncertainty should always be properly labelled and presented along with point estimates for user discretion. Map producers should choose appropriate spatial scales to present model results and refrain from showing high-resolution estimates with extremely low precision.
    \item It is beneficial to use visualization methods that emphasize the probabilistic interpretation of results to identify areas with relatively low or high vaccination coverage. 
    \item Instead of using continuous color scales to show exact point estimates of vaccination coverage, map producers are advised to use our proposed approach to present maps with discrete color scales, which allows comparison and control of the overall precision level of the maps. 
\end{itemize} 

An issue that we briefly mentioned in the introduction but did not discuss at length is the challenge of accounting for the jittering of survey cluster locations in vaccination coverage estimation. Most regression models reply heavily on using geospatial covariates to capture outcome variability and make predictions at locations without survey samples. When survey GPS coordinates are displaced to preserve respondents privacy, the traditional method of overlaying covariate surfaces on point locations is going to match the survey clusters with incorrect covariate values. Unfortunately, this issue has been routinely ignored \cite{takahashi2017geography, mosser2019mapping} or countered with ad-hoc methods \cite{gething2015creating, utazi2018high, utazi2018spatial, utazi2019mapping}. Diligent investigation through simulation studies is needed to understand how much an impact the location displacement could make on model prediction accuracy, especially in the context of covariate modeling. The development of statistically sound methods to overcome this challenge is still an active area of research \cite{wilson2019combining}. This issue again points to not displaying pixel-level maps, at least not with pixels that are smaller than the level of jittering. 

The accurate mapping of vaccination coverage is an important endeavor in the era of the Sustainable Development Goals (SDGs) with the central focus of ``leaving no one behind" \cite{SDGs}. It paves the way towards achieving equity in vaccination by helping us understand the current spatial disparities in attainment of coverage targets. Advances in statistical modeling tools in combination with increased availability of high-resolution geospatial data have enhanced our ability to utilize household surveys to produce vaccination coverage estimates at fine spatial scales. However, it is crucial for us to acknowledge the limitations in the data and methods and use procedures that are appropriate and statistically vetted. Paying attention to details such as survey design and map uncertainty takes care, patience and diligence; but it is definitely worth the effort considering the ultimate goals it helps achieve for our world. 


\section*{Supplementary materials}

Supplementary materials may be found online at the end of the article. \texttt{R} code for the analyses and visualization is available at \url{https://github.com/dq0708/vaxmap}.


\section*{Acknowledgements}

TQD and JW are supported by NIH Grant R01-AI029168.


\section*{Conflict of interest statement}

The authors declare no competing interests.

\bibliography{References}


\section*{Tables and Figures}


\begin{table}[H]
\centering
\begin{adjustwidth}{-2cm}{-0cm}
\resizebox{1.3\columnwidth}{!}{%
\begin{tabular}{cccccccccccc}
\hline \hline
Model Class & Strata Included? & Strata & \multicolumn{2}{c}{Spatial Field} & \multicolumn{2}{c}{Overdispersion} & True Signal & \multicolumn{4}{c}{Model Validation} \\
 &  & Urban $\gamma$ & Range $\rho$ & SE $\sigma_S$ &  $d$ & $\sigma_\delta$ & $\sigma_\epsilon$ & WAIC & Bias & MAE & RMSE\\ \hline \hline
\multirow{2}{*}{Binomial NN} & No Strata &  & \begin{tabular}[c]{@{}c@{}}0.69 \\ (0.47, 1.0)\end{tabular} & \begin{tabular}[c]{@{}c@{}}1.1 \\ (0.96, 1.3)\end{tabular} &  &  &  & 3618  & -0.016 & 0.260 & 0.312\\
 & Strata & \begin{tabular}[c]{@{}c@{}}0.30 \\ (0.10, 0.49)\end{tabular} & \begin{tabular}[c]{@{}c@{}}0.70 \\ (0.47, 1.1)\end{tabular} & \begin{tabular}[c]{@{}c@{}}1.1 \\ (0.97, 1.3)\end{tabular} &  &  &  & 3615 & -0.016 & 0.259 & 0.310\\ \hline 
\multirow{2}{*}{Beta-Binomial OD} & No Strata &  & \begin{tabular}[c]{@{}c@{}}2.9 \\ (1.7, 5.1)\end{tabular} & \begin{tabular}[c]{@{}c@{}}1.6 \\ (1.1, 2.1)\end{tabular} & \begin{tabular}[c]{@{}c@{}}2.9 \\ (2.6, 8.6)\end{tabular} &  &  & 3581  & -0.007 & 0.254 & 0.307 \\
 & Strata & \begin{tabular}[c]{@{}c@{}}0.31 \\ (0.10, 0.51)\end{tabular} & \begin{tabular}[c]{@{}c@{}}3.0 \\ (1.8, 5.3)\end{tabular} & \begin{tabular}[c]{@{}c@{}}1.6 \\ (1.1, 2.1)\end{tabular} & \begin{tabular}[c]{@{}c@{}}2.9 \\ (2.6, 8.6)\end{tabular} &  &  & 3573  & -0.006 & \textbf{0.253} & \textbf{0.305} \\ \hline
\multirow{2}{*}{Lono-Binomial OD} & No Strata & & \begin{tabular}[c]{@{}c@{}}2.7 \\ (1.6, 4.8)\end{tabular} & \begin{tabular}[c]{@{}c@{}}1.4 \\ (1.1, 1.9)\end{tabular} &  & \begin{tabular}[c]{@{}c@{}}0.73 \\ (0.63, 1.9)\end{tabular} &  & 3456  & \textbf{-0.004} & 0.254 & 0.307 \\
 & Strata & \begin{tabular}[c]{@{}c@{}}0.35 \\ (0.13, 0.57)\end{tabular} & \begin{tabular}[c]{@{}c@{}}2.8 \\ (1.7, 5.0)\end{tabular} & \begin{tabular}[c]{@{}c@{}}1.4 \\ (1.1, 1.9)\end{tabular} &  & \begin{tabular}[c]{@{}c@{}}0.73 \\ (0.63, 1.9)\end{tabular} &  & \textbf{3451}   & -0.005 & \textbf{0.253}  & 0.306 \\ \hline 
\multirow{2}{*}{Binomial TS} & No Strata & & \begin{tabular}[c]{@{}c@{}}2.7 \\ (1.6, 4.8)\end{tabular} & \begin{tabular}[c]{@{}c@{}}1.4 \\ (1.1, 1.9)\end{tabular} &  &  & \begin{tabular}[c]{@{}c@{}}0.73 \\ (0.63, 1.9)\end{tabular} & 3456  & -0.006 & 0.254 & 0.307 \\
 & Strata & \begin{tabular}[c]{@{}c@{}}0.35 \\ (0.13, 0.57)\end{tabular} & \begin{tabular}[c]{@{}c@{}}2.8 \\ (1.7, 5.0)\end{tabular} & \begin{tabular}[c]{@{}c@{}}1.4 \\ (1.1, 1.9)\end{tabular} &  &  & \begin{tabular}[c]{@{}c@{}}0.73 \\ (0.63, 1.9)\end{tabular} & \textbf{3451}  & -0.006 & \textbf{0.253} & 0.306 \\ \hline \hline
\end{tabular}%
}
\end{adjustwidth}
\caption{Estimates of parameters and model validation results. Reported are the posterior medians and 2.5\% and 97.5\% quantiles of the regression coefficients and the parameters of the spatial field and cluster-level excess variation. The spatial range parameter $\rho$ is on the longitude-latitude degree scale, which, given the geographical location of Nigeria, equals an average of 111 km per degree. The widely applicable information criterion (WAIC) is calculated based on all the data. The bias, mean absolute error (MAE) and root mean squared error (RMSE) are calculated based on the cross-validation exercise described in the supplementary materials. Bold figures represent the ``best" models according to the relevant criteria.} \label{table1}
\end{table}


\end{document}